\documentstyle[11pt,moriond,epsfig]{article}

\def\Journal#1#2#3#4{{#1} {\bf #2}, #3 (#4)}

\def\NPA{{\em Nucl. Phys.} A}
\def\PLB{{\em Phys. Lett.}  B}
\def\PRL{\em Phys. Rev. Lett.}
\def\PRD{{\em Phys. Rev.} D}
\def\PRC{{\em Phys. Rev.} C}
\def\ZPA{{\em Zeit. Phys.} A}

\def\be{\begin{equation}}
\def\ee{\end{equation}}
\def\bea{\begin{eqnarray}}
\def\eea{\end{eqnarray}}

\def\beq{\begin{equation}}
\def\eeq{\end{equation}}
\begin{document}
\vspace*{4cm}
\title{LOW-MASS DILEPTONS FROM IN-MEDIUM HADRONIC INTERACTIONS}

\author{ R. RAPP}

\address{Department of Physics and Astronomy, State University of New York,\\
Stony Brook, NY 11794-3800, U.S.A.}

\maketitle\abstracts{Medium Effects in low-mass dilepton production in 
ultra-relativistic heavy-ion collisions are investigated using hadronic
models. The rescattering of pions and rho mesons within a hot and dense
hadron gas leads to substantial modifications in the $\pi^+\pi^- \to e^+ e^-$ 
annihilation process, which are shown to be important for understanding 
the recently observed dilepton enhancement at the CERN-SpS. Possible
implications for the nature of the chiral phase transition are outlined.}

\section{Introduction}
A thorough understanding of Quantum Chromodynamics (QCD) has to account 
for the many-body properties of the theory, such as 
its phase diagram. Already at the level of the ground state (vacuum) this 
has proven to be a non-trivial task:    
whereas the fundamental QCD Lagrangian, 
\beq
{\cal L_{\rm QCD}} = \bar q \ (\not\!\!{D}-m_q) \ q -\frac{1}{4} G_{\mu\nu}
G^{\mu\nu} \ ,  
\eeq
exhibits (in the limit of vanishing current quark masses $m_q\to 0$) 
an exact chiral symmetry, the latter is spontaneusly broken in the 
QCD vacuum, leading to, e.g., a nonzero expectation value of the (chiral) 
quark condensate $\langle 0|\bar q_L q_R+\bar q_R q_L|0\rangle$. 
The same inter-quark forces which generate the chiral-symmetry-breakdown 
presumably also govern the structure of the low-lying 
hadronic spectrum, as e.g. indicated by the appearance of (quasi-) 
Goldstone bosons (pions) or the non-observance of parity dubletts (i.e. 
$\pi(140)$$-$$\sigma(400$-$1200)$, $\rho(770)$$-$$a_1(1260)$, 
$N(939)$$-$$N^*(1535)$, 
etc., are seperated in mass by typically $\Delta m\approx 0.5$~GeV).  
However, at sufficiently high temperature and/or density one expects 
chiral symmetry to be restored, within the so-called 'Chiral
Phase Transition'. Thus one concludes that medium 
modifications of hadrons in hot/dense matter are in fact precursors
of chiral symmetry restoration. \\
Experimentally one studies strongly interacting matter in the various  
heavy-ion collision programs at CERN, BNL, GSI, etc.; to probe the 
highest density/temperature 
phases formed in the early stages of central collisions of two heavy nuclei,
electromagnetic observables (photons or dileptons $e^+e^-$, $\mu^+\mu^-$) 
are believed to be particular suitable, since, once produced, they can 
traverse the hadronic fireball without further (strong) interaction. This, 
in principle, provides direct access to the vector meson properties in 
the hadronic medium via their decay modes $V\to e^+e^-, \mu^+\mu^-$ 
($V=\rho, \omega, \phi, J/\Psi, ...$; note, however, that only the $\rho$
meson lifetime of $\tau_\rho$=1.3~fm/c is substantially smaller than the
typical lifetime $\tau_{FB}$=10-20~fm/c of the hadronic fireball). 
In this respect, recent measurements of dilepton invariant mass spectra
at the CERN-SpS have drawn remarkable attention: in the high mass
region, a substantial {\it suppression} of $J/\Psi$ production has been 
observed\cite{Romana,Fereiro}; on the other hand, in the low-mass region, 
a strong {\it increase} in the dilepton yield was found as compared to 
expectations based on hadronic decays after freezeout. Although the  
additional inclusion of $\pi^+\pi^-\to e^+e^-$ annihilation in the hadronic 
fireball substantially increases the dilepton yield, the experimentally
observed excess around invariant masses of $M$$\simeq$0.4~GeV/c$^2$ still
remains unexplained.  \\ 
In this talk I will discuss recent developments in understanding the low-mass 
enhancement in terms of hadronic models and its possible implications for the 
nature of the chiral phase transition. The key object to be calculated is the
so-called electromagnetic (or vector) current correlator (sect.~2), which 
is directly related to the dilepton production rate in hot/dense matter 
as needed for evaluating experimentally observed spectra in heavy-ion
collisions (sect.~3). 


\section{Hadronic Models for the Electromagnetic Current Correlator}


\subsection{Vacuum}\label{subsec:vac}
The electromagnetic (e.m.) current correlation function in free space,  
\beq
\Pi^{vac}_{\mu\nu}(M)=i\int d^4x \ e^{iq\cdot x} \ \langle 0|{\cal T} j_\mu(x) 
j_\nu(0) |0\rangle = (g_{\mu\nu}-\frac{q_\mu q_\nu}{M^2}) \ \Pi^{vac}(M) \ ,
\eeq
is determined by a single scalar function $\Pi^{vac}(M)$ depending on the 
invariant mass $M=\sqrt{q_\mu q^\mu}$ only, the tensor structure being 
fixed by current 
conservation ($q_\mu \Pi^{\mu\nu}$=0). Invoking the well-established 
vector dominance model (VDM), the hadronic e.m. current is saturated by 
the low-lying vector mesons, i.e. 
$j^\mu$=$j^\mu_\rho$+$j^\mu_\omega$+$j^\mu_\phi$.   
For what follows, the $\rho$ meson part (isospin $I$=1) will be the most 
relevant one; it is given by  
\beq
Im \Pi_{I=1}^{vac}(M)=\frac{Im\Sigma_\rho^{vac}(M)}{g_\rho^2} \ |F_\pi(M)|^2 
=\left( \frac{(m_\rho^{(0)})^2}{g_\rho}\right)^2 Im D_\rho^{vac}(M) \
\eeq
with $F_\pi$: pion electromagnetic formfactor, and 
$D_\rho^{vac}(M)=[M^2-(m_\rho^{(0)})^2-\Sigma_\rho^{vac}(M)]^{-1}$: free
$\rho$ propagator
consisting of a bare $\rho$ of mass $m_\rho^{(0)}$, dressed with 
intermediate two-pion states ('pion cloud') represented by the $\rho$ 
selfenergy $\Sigma_{\rho\pi\pi}^{vac}$. This model 
is consistent with experimental data on both $F_\pi$ and p-wave $\pi\pi$ 
scattering (proceeding through the $\rho$ resonance)~\cite{RCW}. 


\subsection{Medium Modifications}
In hadronic matter of temperature $T$ and baryo-chemical potential $\mu_B$,
the e.m. correlator reads  
\bea
\Pi_{\mu\nu}(q_0,q;\mu_B,T) & = & i\int d^4x \ e^{iq\cdot x} \  
Tr[e^{(H-\mu_BN)/T} {\cal T} j_\mu(x) j_\nu(0)] / {\cal Z}
\nonumber\\
& = & \Pi^L(q_0,q;\mu_B,T) \ P^L_{\mu\nu}+\Pi^T(q_0,q;\mu_B,T) \ P^T_{\mu\nu} 
\eea 
where the breaking of Lorentz invariance implies the existence of two 
independent modes (longitudinal and transverse), depending separately
on energy $q_0$ and 3-momentum modulus $q$.  \\
Various approaches have been pursued in the literature to assess medium 
effects in $\Pi_{\mu\nu}$; the most spectacular one is the so-called 
Brown-Rho scaling conjecture~\cite{BR91}, where in a mean field type 
picture the vector meson masses are directly linked to the reduction 
of the chiral condensate towards the chiral phase transition.
Such a scenario leads to good agreement with the observed dilepton spectra
at the CERN-SpS~\cite{LKBS}. 
More conventional ones include a 'chiral reduction formalism'~\cite{SYZ}, 
which ties chiral Ward identities with experimental information,  
or a combination of chiral SU(3) Lagrangians with VDM~\cite{KKW}; in both 
these frameworks medium effects are based on a low-density expansion.    
We here discuss a slightly different approach, where phenomenological 
information on in-medium $\pi N$ and $\rho N$ interactions is employed
within the VDM. Restricting ourselves first to the case of nuclear matter
at zero temperature two types of medium effects in the $\rho$ propagator
as introduced in sect.~\ref{subsec:vac} arise: 
\begin{itemize}  
\item[(i)] renormalization of the pion cloud through pion interactions 
with the surrounding nucleons~\cite{rhopp,RCW}; as is well 
known from the analysis of 
pion nuclear optical potentials, the dominant contributions stem from
p-wave nucleon-nucleonhole and delta-nucleonhole excitations including
short-range correlation effects (parametrized by Migdal parameters $g'$); 
\item[(ii)] direct scattering of the $\rho$ on surrounding nucleons, which,
in analogy to (i), we assume to be dominated by s-channel baryon ($B$) 
resonance graphs~\cite{FrPi,RCW,PPLLM,RUBW}; the coupling constants of the 
corresponding $\rho BN$ vertex can be estimated from the free partial 
decay width $\Gamma^0_{B\to\rho N}$; they will be quantitatively fixed 
in the following section. The resonances included are listed in 
table~\ref{tab:res}; note that the finite width of the $\rho$ allows 
decays of the type $B\to\rho N \to\pi\pi N$ well below the rho-nucleon 
threshold of $m_N+m_\rho$.    
\end{itemize} 

\begin{table}[h]
\caption{Baryon resonances with substantial coupling to $\rho N$ that are
included in our calculation; the second row gives the spin-parity $J^P$ 
of the resonance, which
determines whether the $\rho N$ coupling is of $p$-wave (for $P=+$) or 
$s$-/$d$-wave (for $P=-$) type; the third row gives the partial decay width 
(in MeV) for $B\to\rho N$.  \label{tab:res}}
\vspace{0.3cm}
\begin{center}
\begin{tabular}{|c|cccccccc|}
\hline
 $B^P$ & $N(939)^+$  & $\Delta(1232)$ & $N(1440)$ & $N(1520)$ & 
$\Delta(1620)$ & $\Delta(1700)$ & $N(1720)$ & $\Delta(1905)$ \\
\hline
$J^P$ & $\frac{1}{2}^+$ & $\frac{3}{2}^+$ & $\frac{1}{2}^+$ & $\frac{3}{2}^-$ & 
$\frac{1}{2}^-$ & $\frac{3}{2}^-$ & $\frac{3}{2}^+$ & $\frac{5}{2}^+$ \\
 $\Gamma_{\rho N}^0$ & $-$ & $-$ & $\sim 10$ & $\sim 25$ & $\sim 15$ & 
 $\sim 100$&  $>100$ & $>200$    \\ \hline
\end{tabular}
\end{center}
\end{table}

\begin{figure}
\psfig{figure=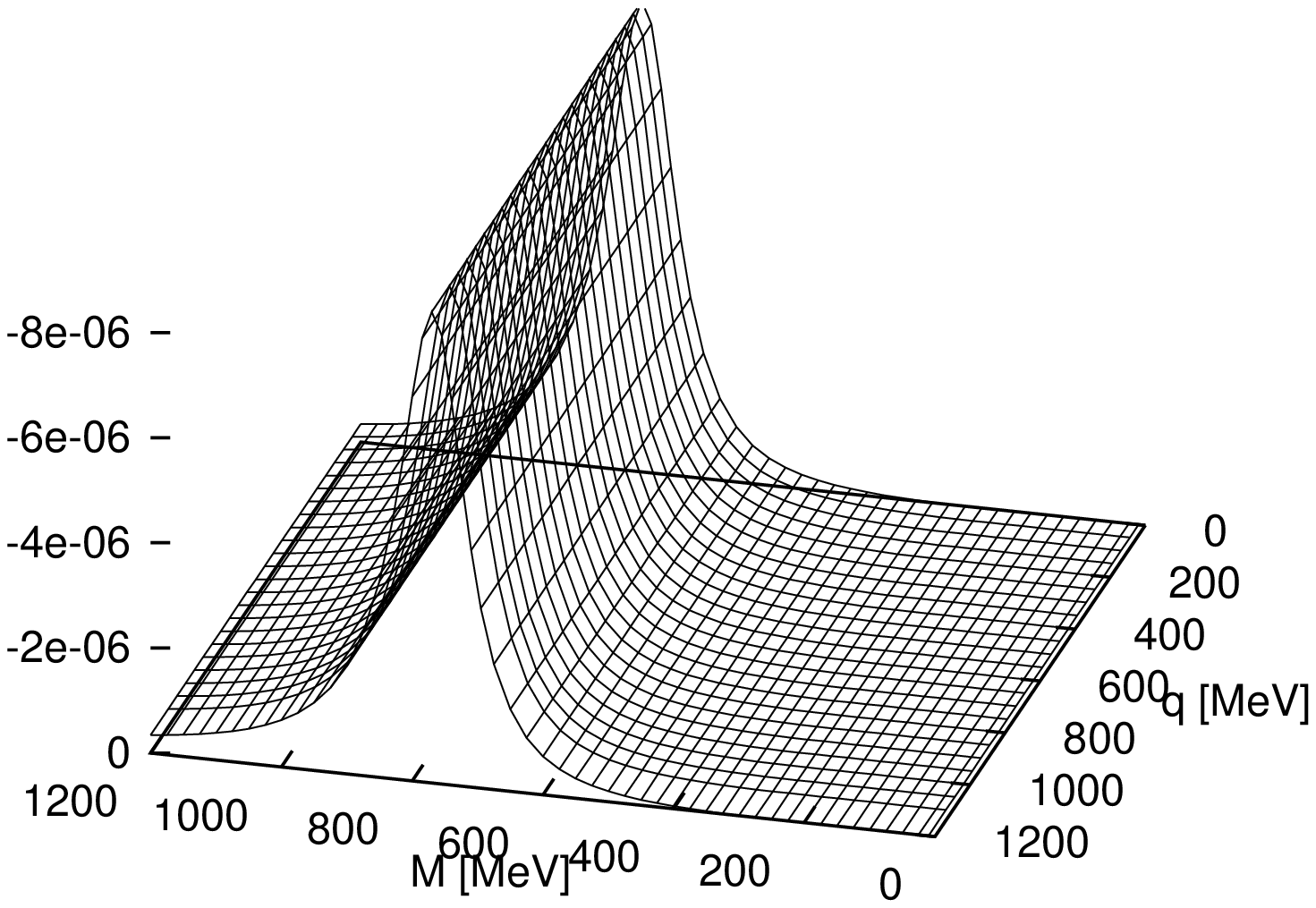,height=3.1in,width=2.5in,angle=-90}
\psfig{figure=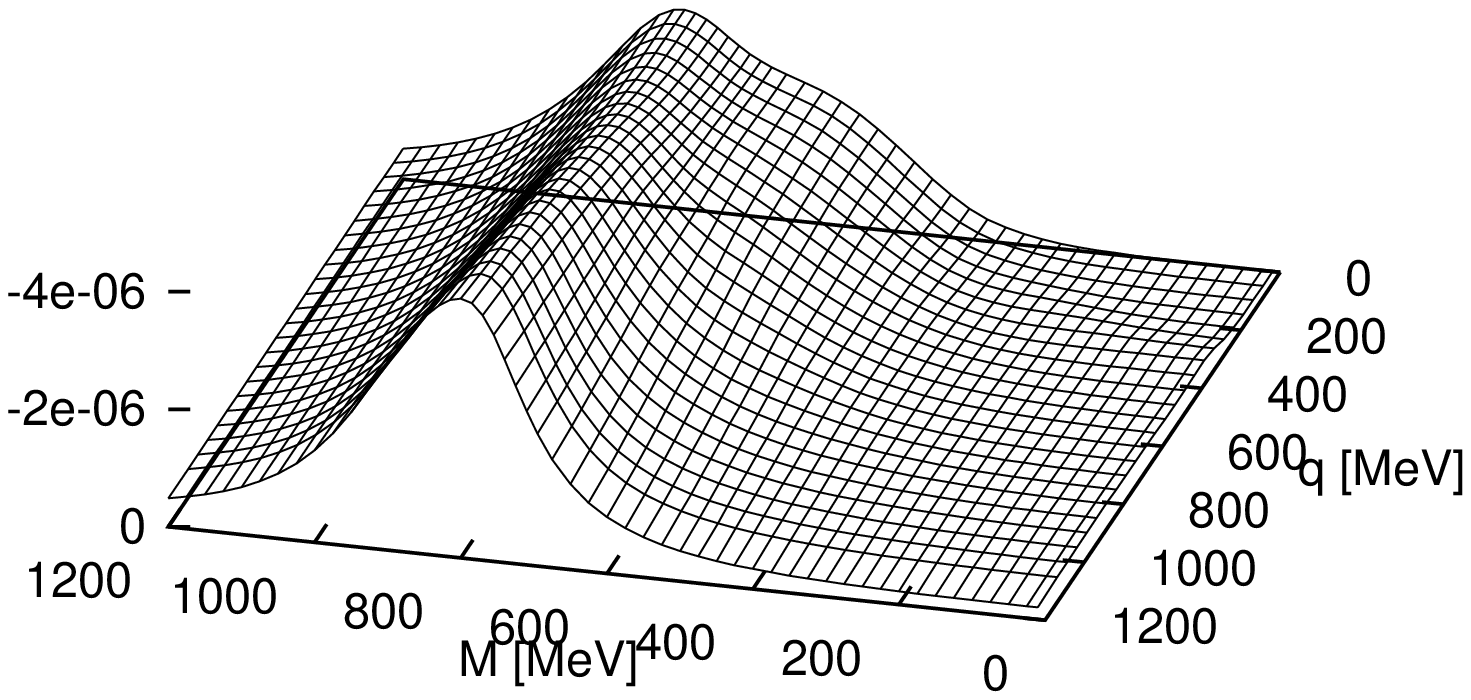,height=3.1in,width=2.5in,angle=-90}
\caption{$\rho$-meson spectral function versus invariant mass $M$ and 
3-momentum $q$ in vacuum (left panel) and in normal
nuclear matter (right panel).
\label{fig:rhoprop}}
\end{figure}

Thus the longitudinal and transverse parts of the $\rho$ propagator at nuclear 
density $\rho_N$ become 
\beq
D_\rho^{L,T}(q_0,q;\rho_N)=\left[ q_0^2-q^2-(m_\rho^{(0)})^2-
\Sigma_{\rho\pi\pi}^{L,T}(q_0,q;\rho_N)-
\Sigma_{\rho BN}^{L,T}(q_0,q;\rho_N) \right]^{-1} \ . 
\label{drhomed} 
\eeq
The spin-averaged $\rho$ spectral function 
$Im D_\rho=\frac{1}{3}(ImD_\rho^L+2ImD_\rho^T)$ is shown in 
fig.~\ref{fig:rhoprop}:
already at normal nuclear matter density $\rho_N=\rho_0=0.16$~fm$^{-3}$  
one observes a strong broadening of the resonance (right panel)
as compared to free space (left panel). 


\subsection{Model Constraints from Experimental Data} 
\begin{figure}[h]
\psfig{figure=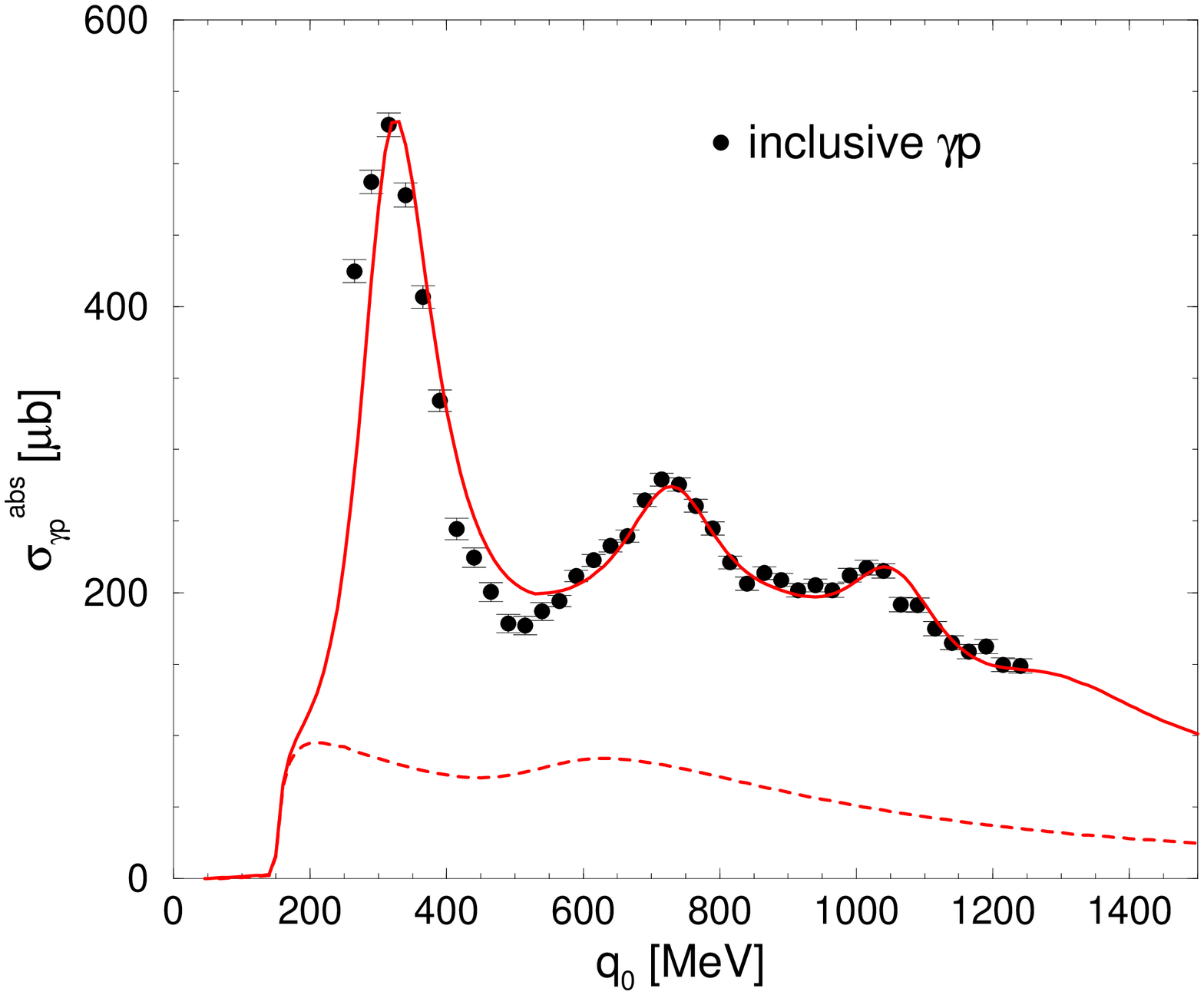,height=3.1in,width=2.5in,angle=-90}
\psfig{figure=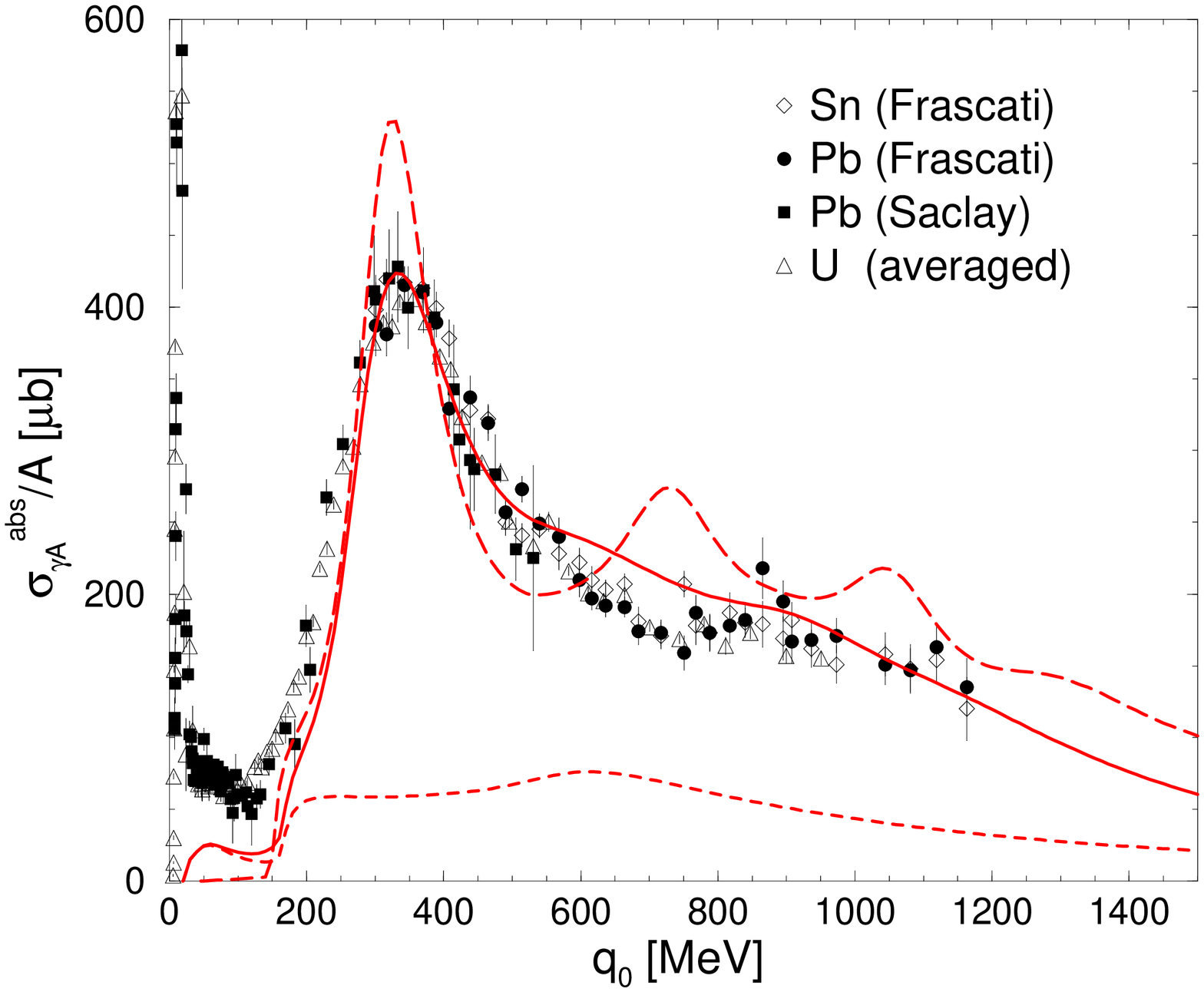,height=3.1in,width=2.5in,angle=-90}
\caption{The photo absorption spectrum on the proton\protect\cite{photop}
(left panel) and on nuclei~\protect\cite{photoA} (right panel). The solid  
lines are the full results in the low-density limit (left panel) and for  
$\rho_N=0.8\rho_0$ (right panel), and short-dashed lines indicate the 
non-resonant background contributions.}
\label{fig:photo}
\end{figure}

To increase the reliability of the model when calculating dilepton 
production in heavy-ion collisions it is important to check   
consistency with other related data. Obviously, 
photoabsorption processes represent the $M\to 0$ limit
of the (timlike) dilepton regime and therefore provide valuable 
constraints~\cite{RUBW}. 
Within the VDM the total photoabsorption cross section on a single nucleon
or on nuclei (normalized to the number of nucleons $A$) 
can be written as
\beq 
\frac{\sigma_\gamma^{abs}}{A}=\frac{4\pi\alpha}{g_\rho^2 q_0} \ \frac{1}{\rho_N}
 \ \bar F (q_0=q;\rho_N) \ , 
\label{sigabs}
\eeq
where $\bar F$ is essentially proportional to the imaginary part of the 
transverse $\rho$ propagator, Eq.~(\ref{drhomed}). It includes direct 
resonance formation (encoded in $\Sigma_{\rho BN}$) and interactions
via the pion cloud (the so-called meson exchange currents or 'background' 
contributions encoded in $\Sigma_{\rho\pi\pi}$), see the previous section.
In the low-density limit, $\rho_N\to 0$, Eq.~(\ref{sigabs}) 
corresponds to the absorption process on a single nucleon, 
the results for which are displayed in the left panel of 
fig.~\ref{fig:photo}: the spectrum is dominated by the resonance 
contributions with the background constituting about 20-30\% of the 
strength. For finite nuclei, most of the resonance structure disappears, 
indicating an in-medium broadening of the higher baryon resonances as 
well; the similarity of the data over a wide range of atomic numbers 
suggests that nuclear structure effects are not important, thus justifying 
to perform our calculation for infinite nuclear matter at an average density
of $\rho_N=0.8\rho_0$ (full curve in the right panel of fig.~\ref{fig:photo}). 
Apparently, the lowest-order-in-density result (long-dashed curve) does not
provide a satisfactory description of the nucleus data. 

\noindent
Further model constraints can be obtained, e.g., from the analysis 
of $\pi N \to \rho N$ production~\cite{Friman}, which is in fact  
dominated by 'background' contributions. This imposes stringent 
constraints on the hadronic formfactor at the $\pi NN$ vertex (included 
in the results shown in fig.~\ref{fig:photo}~\cite{RBRW}).  


\section{Dilepton Production in URHIC's} 

\begin{figure}
\psfig{figure=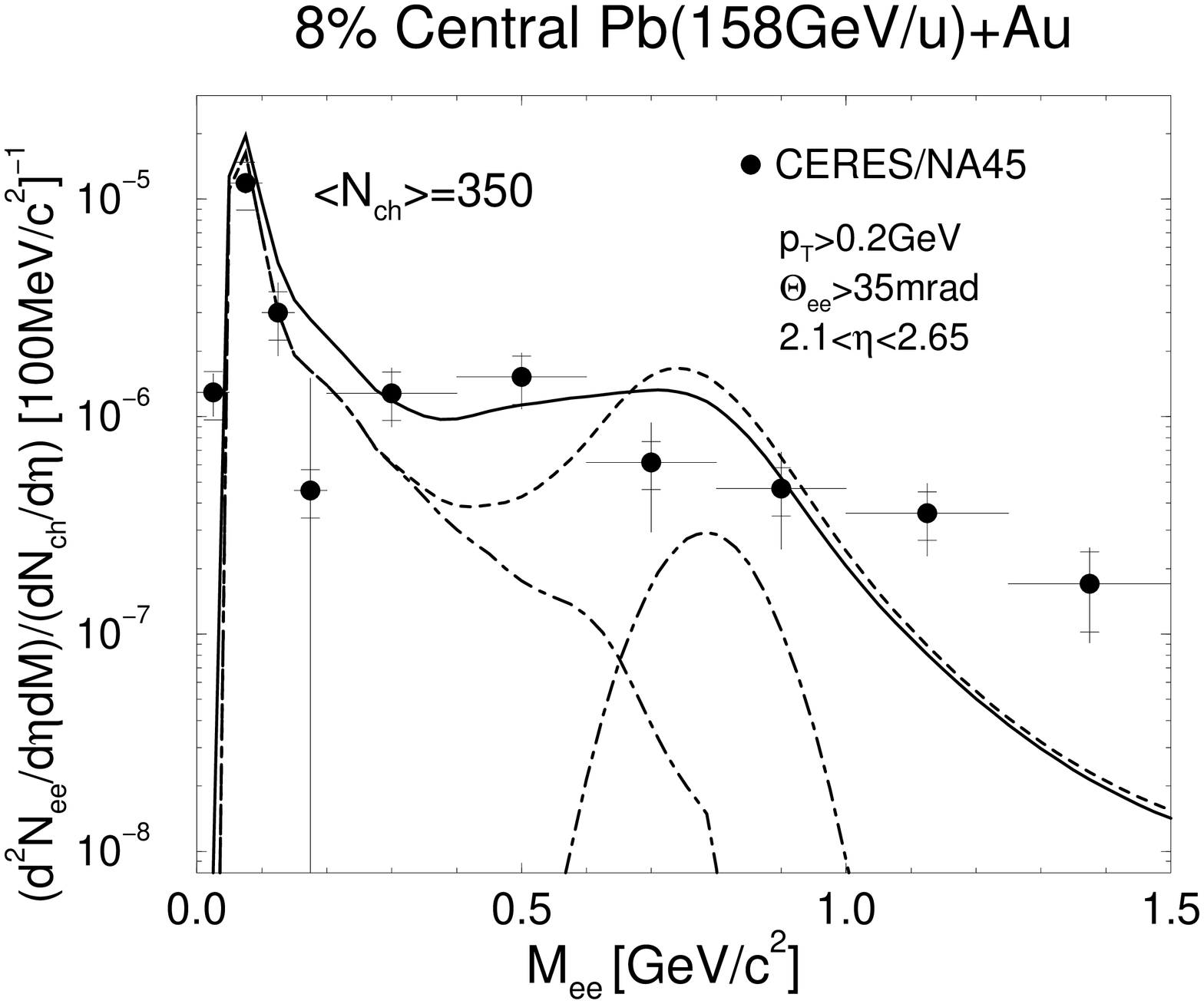,height=3.1in,width=2.5in,angle=-90}
\psfig{figure=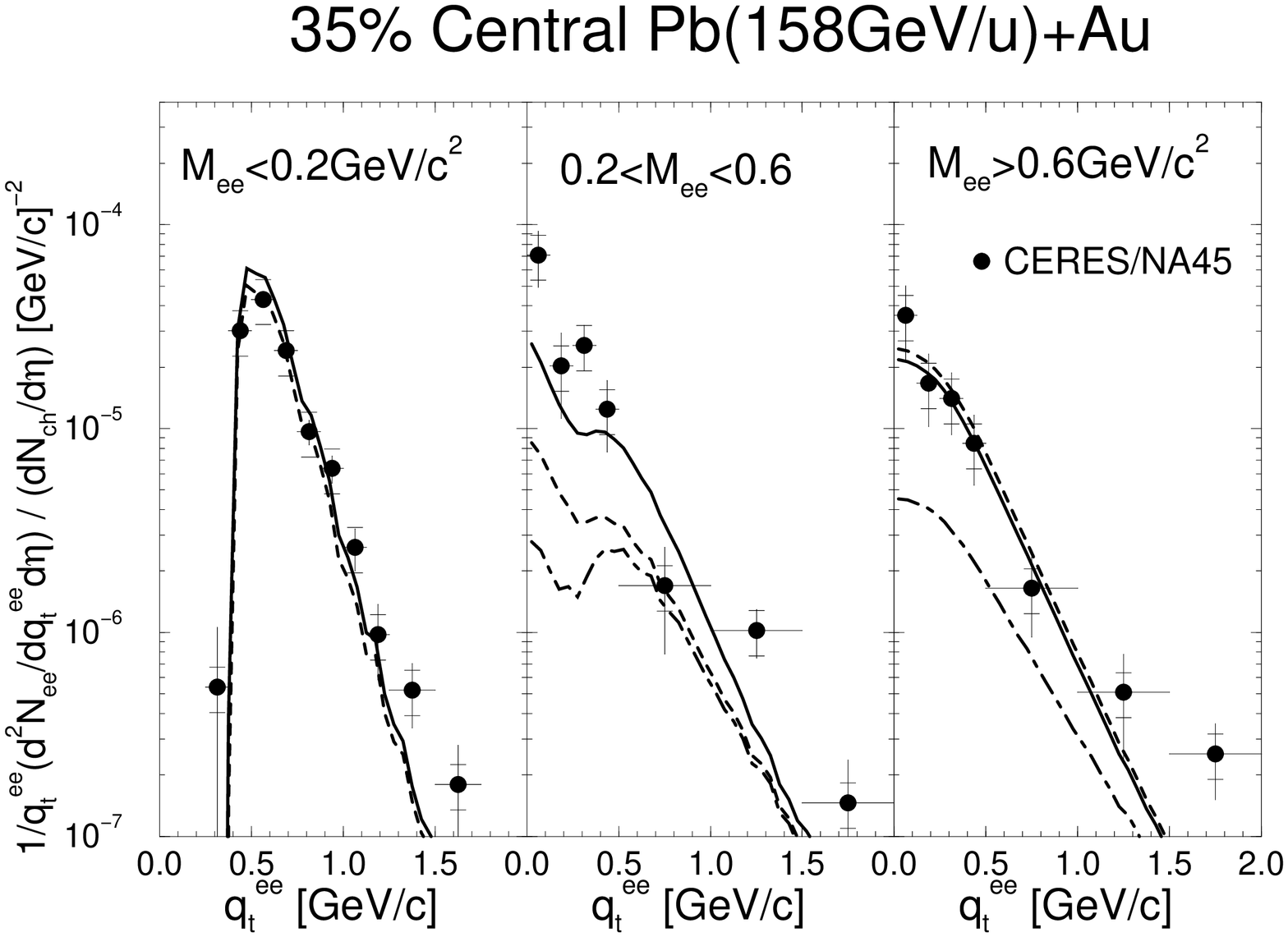,height=3.1in,width=2.5in,angle=-90}
\caption{Dilepton invariant mass (left panel) and
transverse momentum (right panel)
spectra in Pb+Au collisions at CERN-SpS
energies~\protect\cite{Glaessel}. The dashed-dotted lines are calculated
from hadron decays after freezeout; adding the contribution
from $\pi^+\pi^-$ annihilation in the hadronic fireball, one
obtains the dashed lines (when using the free $\rho$ propagator)
or the full lines (when using the in-medium $\rho$ propagator).
\label{fig:dlspec}}
\end{figure}

In URHIC's at CERN-SpS energies (160-200GeV/u) several hundred secondary 
particles
are produced (mostly pions), together with substantial temperatures of the
hadronic fireball of about $T\approx120$-200~MeV, which has to be 
accounted for in a realistic application of the $\rho$ propagator. 
To this end we 
use retarded (finite temperature) selfenergies for both the pion cloud
and the baryon resonance contributions, and include direct interactions
of the $\rho$ meson with pions and kaons from the heat bath via 
$a_1(1260)$ and $K_1(1270)$ resonance formation. \\
Using VDM, the $e^+e^-$ production rate from $\pi^+ \pi^-$ annihilation 
during the interaction phase of the fireball is related to the imaginary 
part of the $\rho$ propagator as 
\beq
\frac{dN_{\pi^+\pi^-\to e^+e^-}}{d^4x \ d^4q}=
\frac{\alpha^2(m_\rho^{(0)})^4}{\pi^3 g_\rho^2} \ f^\rho(q_0;T) \  
\frac{1}{M^2} \ \frac{(-1)}{3} \ (Im D_\rho^L+2 \ Im D_\rho^T) \ , 
\eeq
where $f^\rho$ denotes the thermal (bose-) occupation factor for $\rho$ 
mesons. This rate is to be integrated over the space-time history of
a given heavy-ion reaction, which can be modelled by e.g. 
hydrodynamic~\cite{hydro} or transport~\cite{CBRW} simulations, or simply by 
an expanding fireball. The latter approach is determined by a temperature
and density evolution (with isotropic volume $V_{FB}(t)$), which, 
after including experimental acceptance, 
leads to a dilepton spectrum according to
\beq
\frac{dN_{\pi^+\pi^-\to e^+e^-}}{dM \ d\eta} = \int\limits_{0}^{t_{fo}}
dt \ V_{FB}(t) \int d^3q \ 
\frac{dN_{\pi^+\pi^-\to e^+e^-}}{dt \ d^4q}(q_0,q;\rho_B(t),T(t)) \ 
Acc(q_0,\vec q) \ .   
\label{dndm}
\eeq    
Further contributions arise from hadron decays {\it after} freezeout 
(such as Dalitz decays $\pi^0,\eta\to \gamma e^+e^-$, $\omega\to \pi^0e^+e^-$
or $\omega,\rho^0\to e^+e^-$), taken from the CERES 
collaboration~\cite{Glaessel} (dashed-dotted lines in fig.~\ref{fig:dlspec}), 
which have to be added to the final spectra. 
Our final results are compared to  CERES/NA45 data for Pb(158GeV/u)+Au 
collisions~\cite{Glaessel} in 
fig.~\ref{fig:dlspec}: when using the free $\rho$ spectral function in 
Eq.~(\ref{dndm}), the low-mass enhancement around $M$$\simeq$0.4~GeV cannot
be explained (dashed curves); however, when including the in-medium 
effects due to hadronic rescattering as discussed above, reasonable 
agreement is obtained (full curves) with both the invariant mass (left panel) 
and transverse momentum spectra (right panel). Note that the major part
of the enhancement for 0.2~GeV$<$$M_{ee}$$<$0.6~GeV is correctly ascribed 
to rather small pair momenta $q_t$$\leq$0.7~GeV. This might in fact resolve
the question why the $\mu^+\mu^-$ spectra measured by the NA50 collaboration
show a much less pronounced excess at low $M_{\mu\mu}$: their transverse 
momentum cut of $p_t$$>$1~GeV on the single muon tracks will eliminate most of 
enhancement generated within our model.

\section{Summary and Conclusions} 
We have shown that in-medium hadronic interactions generate a substantial
excess of low-mass dileptons, which can essentially account for the 
experimentally observed enhancement in heavy-ion reactions at CERN-SpS 
energies.  
The question whether the 'melting' of the $\rho$ resonance as found in our 
analysis might signal (partial) restoration of chiral symmetry in strongly
interacting matter remains open. 
Further insight could be gained, e.g., by investigating 
the in-medium properties of the $a_1(1260)$ meson, which, towards the chiral
phase transition, has to become degenerate with its chiral partner, the $\rho$
meson. Experimentally, however, this will be difficult to assess.

\section*{Acknowledgments}
It is a pleasure to thank J. Wambach for fruitful collaboration from  
which many of the results presented here emerged.  
I furthermore thank G.E. Brown, M. Buballa, W. Cassing, A. Drees and 
E. Shuryak for 
useful discussions.  This work was supported in part by 
the A.-v.-Humboldt foundation (within a Feodor-Lynen fellowship) and 
the US-DOE under grant no. DE-FG02-88ER40388.

\end{document}